\documentclass[epj]{webofc}
\usepackage[varg]{txfonts}   
\woctitle{XLVI International Symposium on Multiparticle Dynamics}
\begin{document}
\title{Prompt atmospheric neutrino flux from the various QCD models  }
%
%

\author{Yu Seon Jeong\inst{1}\fnsep\thanks{\email{ysjeong@kisti.re.kr}}
\and      Atri Bhattacharya\inst{2}
\and     Rikard Enberg\inst{3}
\and     C. S. Kim\inst{4}
\and     Mary Hall Reno\inst{5}
\and \\  Ina Sarcevic\inst{6}
\and     Anna Stasto\inst{7}
}

\institute{  National Institute of Supercomputing and Networking, KISTI, Daejeon 34141, Korea 
\and         Space sciences, Technologies and Astrophysics Research (STAR) Institute, Universit\'{e} de Li\`{e}ge, B\^{a}t.~B5a, 4000 Li\`{e}ge, Belgium 
\and         Department of Physics and Astronomy, Uppsala University, Uppsala, Sweden
\and         Department of Physics and IPAP, Yonsei University, Seoul 03722, Korea 
\and	        Department of Physics and Astronomy, University of Iowa, Iowa City, Iowa 52242 
\and	        Department of Astronomy, University  of Arizona, 933 N.\ Cherry Ave., Tucson, AZ 85721 
\and	        Department of Physics, The Pennsylvania State University, University Park, PA 16802
          }

\abstract{%
We evaluate the prompt atmospheric neutrino flux using the different QCD models for heavy quark production including the $b$ quark contribution.
We include the nuclear correction and find it reduces the fluxes by $10 \% - 50\%$ according to the models.  Our heavy quark results are compared with experimental data from RHIC, LHC and LHCb.
}
\maketitle
\section{Introduction}
\label{intro}
The IceCube collaboration reported that they have observed 54 high energy events for the neutrino candidate events
and rejected a pure atmospheric origin scenario with about 7$\sigma$ \cite{icecube}.  
For astrophysical neutrinos, the main background is the atmospheric neutrinos produced from the interaction of cosmic rays with air nuclei.  
In estimating the atmospheric neutrino flux, the most important factor is the cross sections for production of quarks fragmented into hadrons,
which subsequently decay and generate neutrinos.
Above $\sim$1 PeV of energies, atmospheric neutrinos from the decay of heavy hadrons (called prompt neutrinos)
are more important than those from the $\pi$ or $K$ decay (conventional neutrinos).

In this work, we investigate the effect of the QCD models for heavy quark production to the prompt atmospheric neutrino flux. 
We evaluated the prompt flux using three different models, 
next-to-leading order (NLO) perturbative QCD, the dipole model and $k_T$ factorization, and we include $B$ hadron contributions.
To investigate the uncertainties, we vary the QCD scales so that the total cross sections for $q\bar{q}$ pair production 
are consistent with the data from RHIC and LHC experiments.
We find that the scale ranges used here for the cross sections are acceptable by comparing the rapidity distributions of the differential cross sections 
with the corresponding LHCb data.
For other uncertainty factors, we include the nuclear corrections 
and used the modern cosmic ray fluxes parameterized by their composition and source component.   
In this contribution, we present the prompt neutrino fluxes evaluated from the different models for only one cosmic ray spectrum. 
Our comprehensive results can be found in \cite{bejkrss}.

\section{QCD models for heavy quark production}
\label{sec-2}

\begin{figure}
\centering
\sidecaption
 \includegraphics[width=7cm,clip]{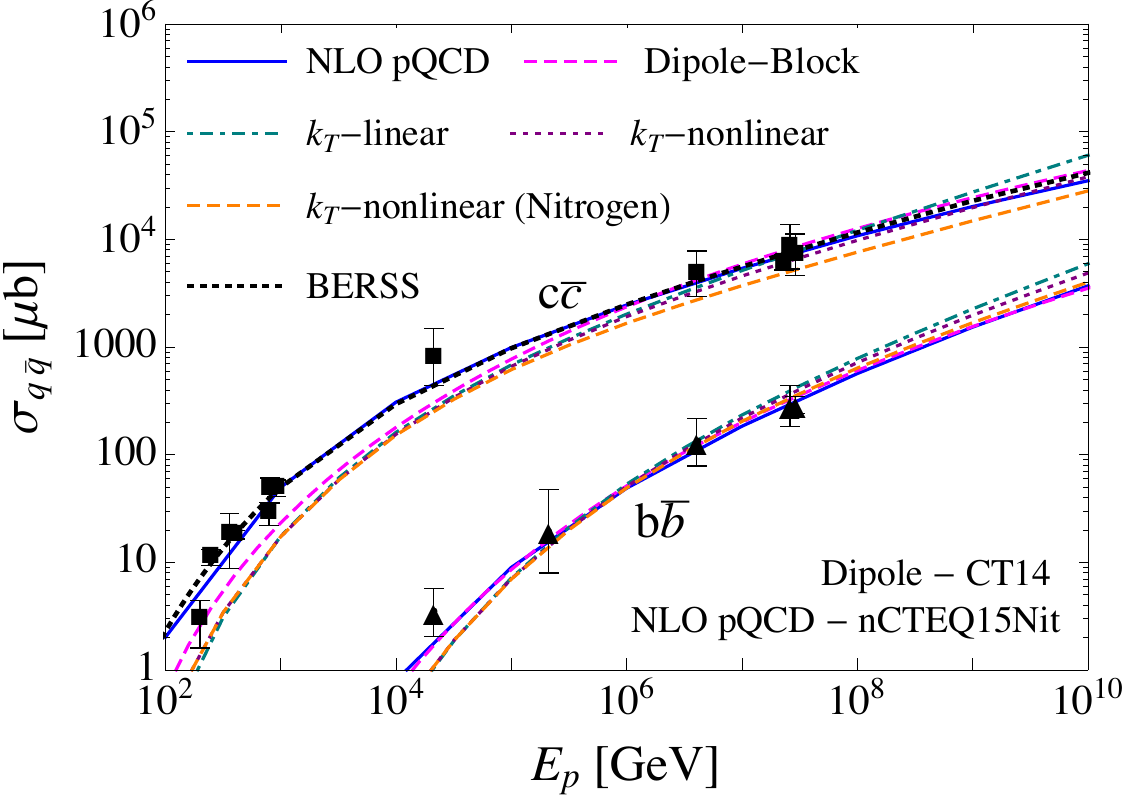}
\caption{ Total cross section for $c\bar{c}$ and $b\bar{b}$ production as a function of the incident proton energy. 
The results are presented for the different models described in section \ref {sec-2}.
For comparison, the previous evaluation at NLO in pQCD approach is presented as BERSS \cite{berss}.}
\label{fig-hqx}     
\end{figure}

\subsection{Perturbative QCD}
\label{sec-pQCD}
The standard method to evaluate heavy quark production is to use the parton model in perturbative QCD (pQCD) with a collinear approximation.
In the pQCD framework, the differential cross section for $q \bar{q}$ pair production is given by 
\begin{equation}
\frac{d \sigma (pp \rightarrow q \bar{q} X) }{d x_F} = 
 \int \frac{d M^2_{q \bar{q}}}{(x_1+x_2)s}
 \sigma_{f_a f_b \rightarrow q\bar{q}}(\hat{s})
 f_a(x_1, M_F^2) f_b(x_2, M_F^2)
\end{equation}
with the parton distribution function (PDF) $f_{a (b)}$, their momentum fraction $x_{1 (2)}$ and the Feynman variable $x_F$. 
Momentum fractions can be expressed as 
$x_{1,2} =\bigl( (x_F^2 + 4 M^2_{q\bar{q}}/s)^{1/2}\pm x_F \bigr) /2$.  For the forward scattering that dominates the prompt flux, 
the partons from the incoming cosmic ray hadrons have the large momentum fraction $x_1$ while those from the target nucleus side have the small $x_2$. 

We compute the heavy quark production cross section at next-to-leading order (NLO) to evaluate the prompt neutrino flux
including the nuclear corrections. 
Nuclear corrections are included through the recent nuclear PDF, nCTEQ15 \cite{ncteq}.
For the factorization ($M_F$) and renormalization ($M_R$) scales, we use scales
proportional to the transverse mass $m_T$,  $(2.1, 1.6)m_T$ for the central values, as experimentally constrained by the RHIC and LHC data \cite{nvf}.  

At the high energies where the prompt flux dominates, gluon fusion to $q\bar{q}$ is the main process, 
hence the essential factor is  the poorly constrained gluon distribution at small-$x$. Therefore we investigate other possible approaches to the extrapolation with  the resummation of large logs of $x$.

\subsection{Dipole model}
\label{sec-3}
One of the alternative approaches is a color dipole model, which can include the parton saturation effect through the so-called dipole cross section. 
In the dipole model, the heavy quarks are produced through two sequential processes: a fluctuation of incoming gluon into $q \bar{q}$ pair (color dipole), 
and the interaction of the color dipoles with target nuclei. 
The dipole cross section is given by 
\begin{align}
\sigma^{g p \to q\bar q X}(x,M_R,Q^2) = \int dz \, d^2\vec r \,
|\Psi^q_g(z,\vec r,M_R,Q^2)|^2 \sigma_{d}(x,\vec r)\ 
\label{eq:dipolexsec}
\end{align}
with the wave function squared ($|\Psi^q_g|^2$) for a gluon fluctuation to a color dipole 
and the dipole cross section ($\sigma_d$) for the interaction between the dipole and the target. 
The $x_F$ differential cross section for heavy quark pair production in $pp$ collision can be written as 
\begin{equation}
\frac{d\sigma (pp\to q\bar{q}X)}{dx_F}\simeq \frac{x_1}{x_1+x_2}
 g(x_1,M_F)\sigma^{gp\to q\bar{q}X}
(x_2,M_R,Q^2=0)\ .
\end{equation}

\begin{figure}
\centering
 \includegraphics[width=4.53cm,clip]{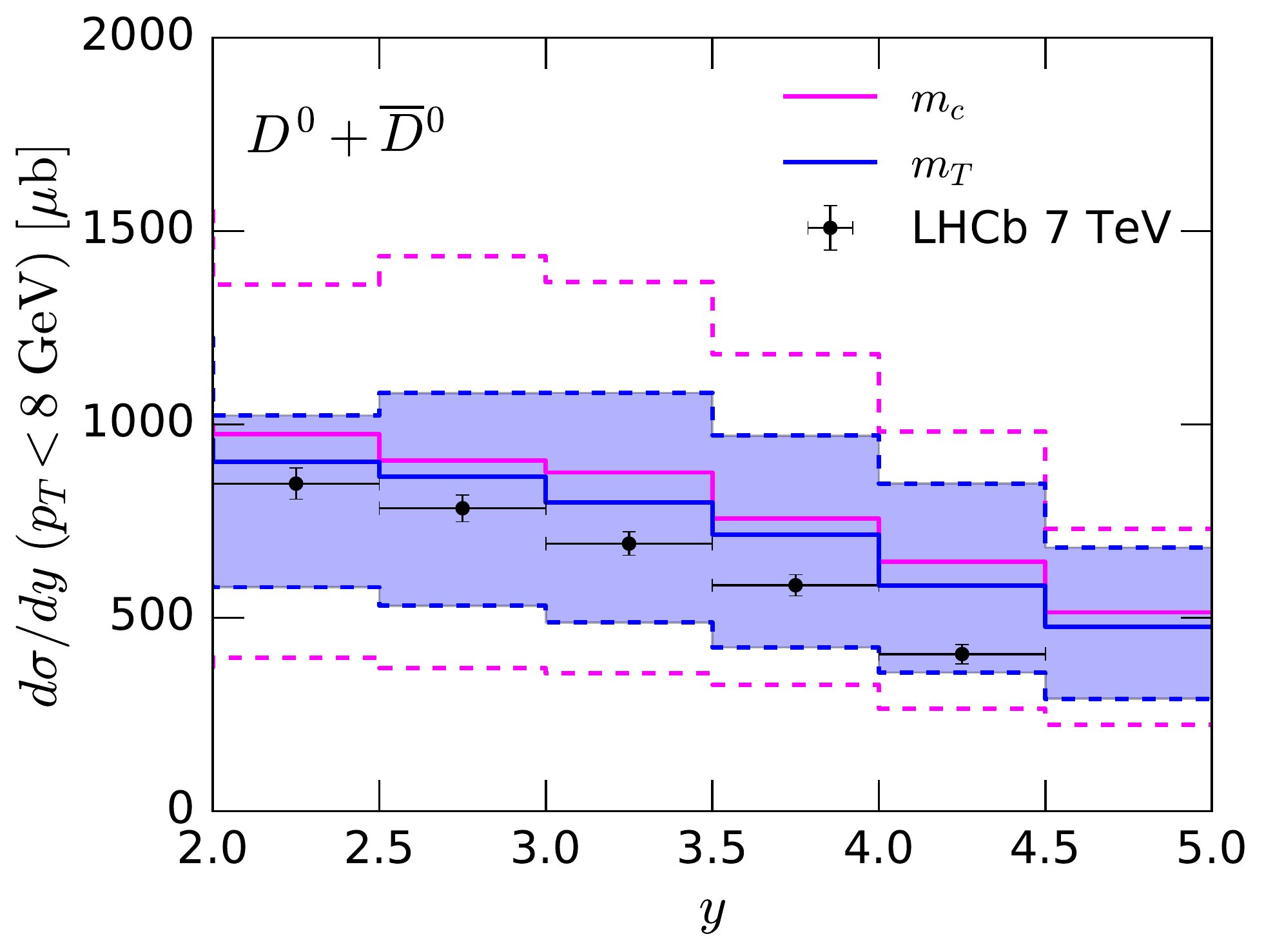}
 \includegraphics[width=4.75cm,clip]{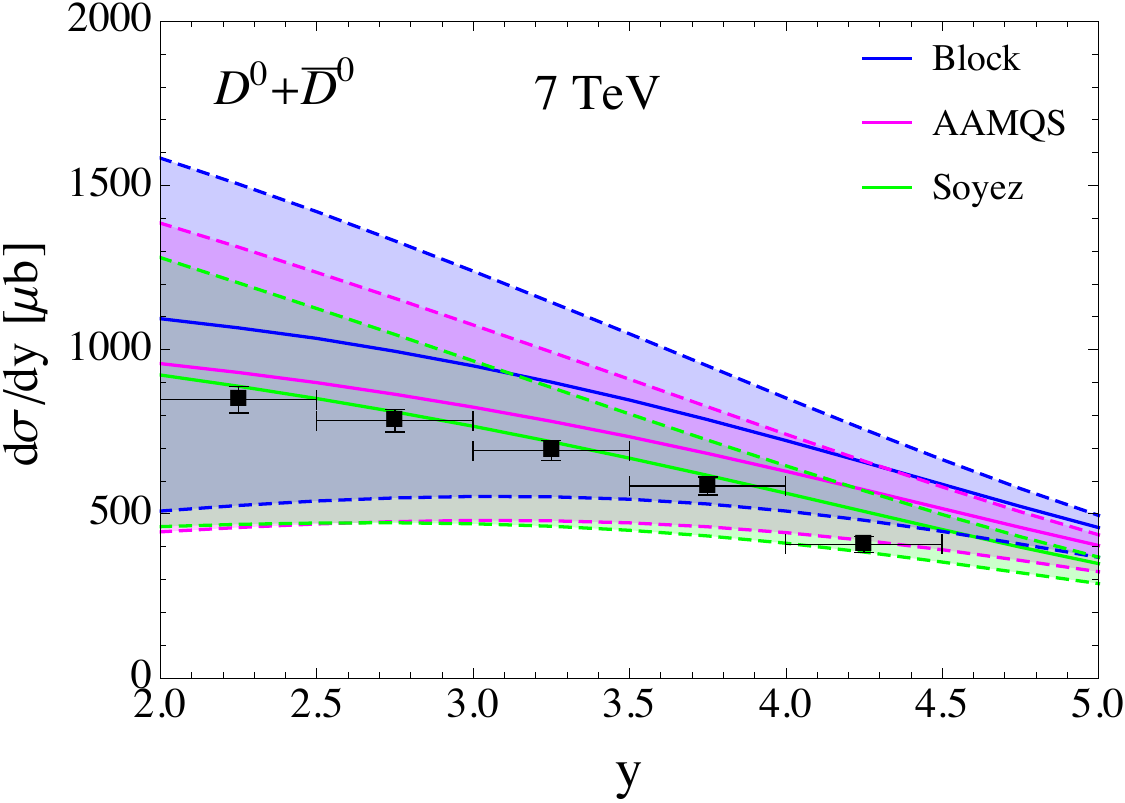}
 \includegraphics[width=4.75cm,clip]{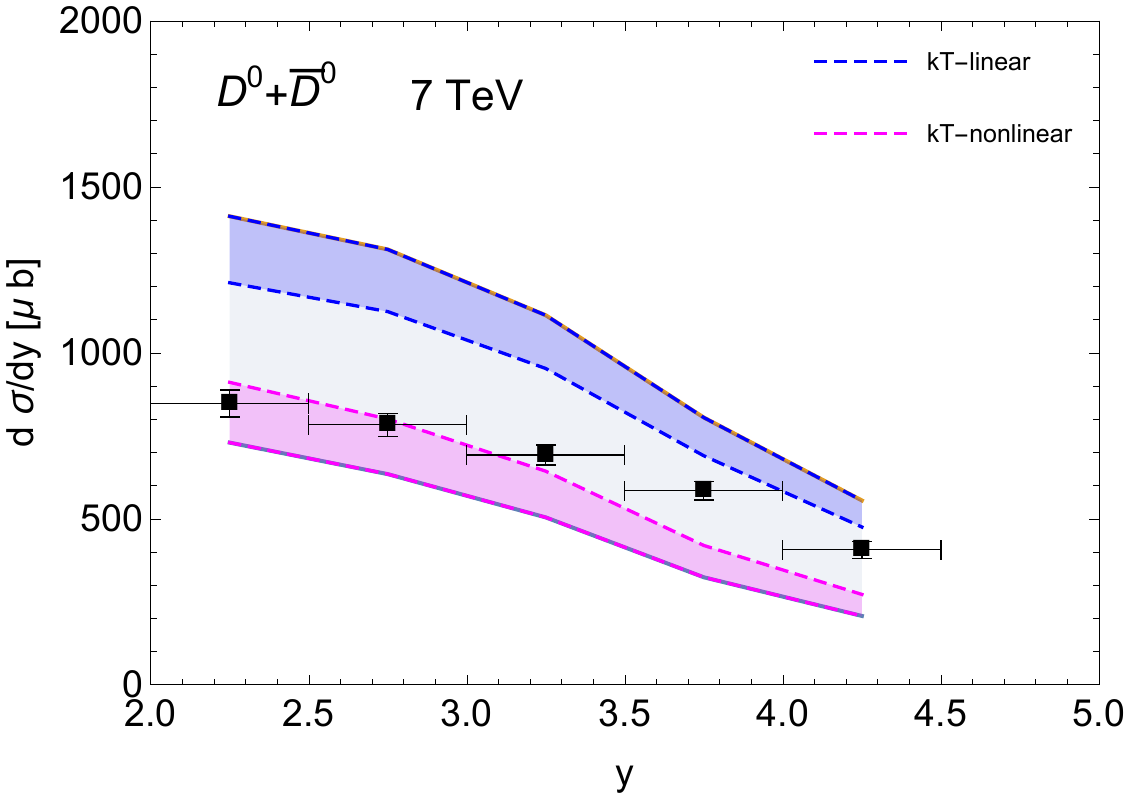}
 \includegraphics[width=4.53cm,clip]{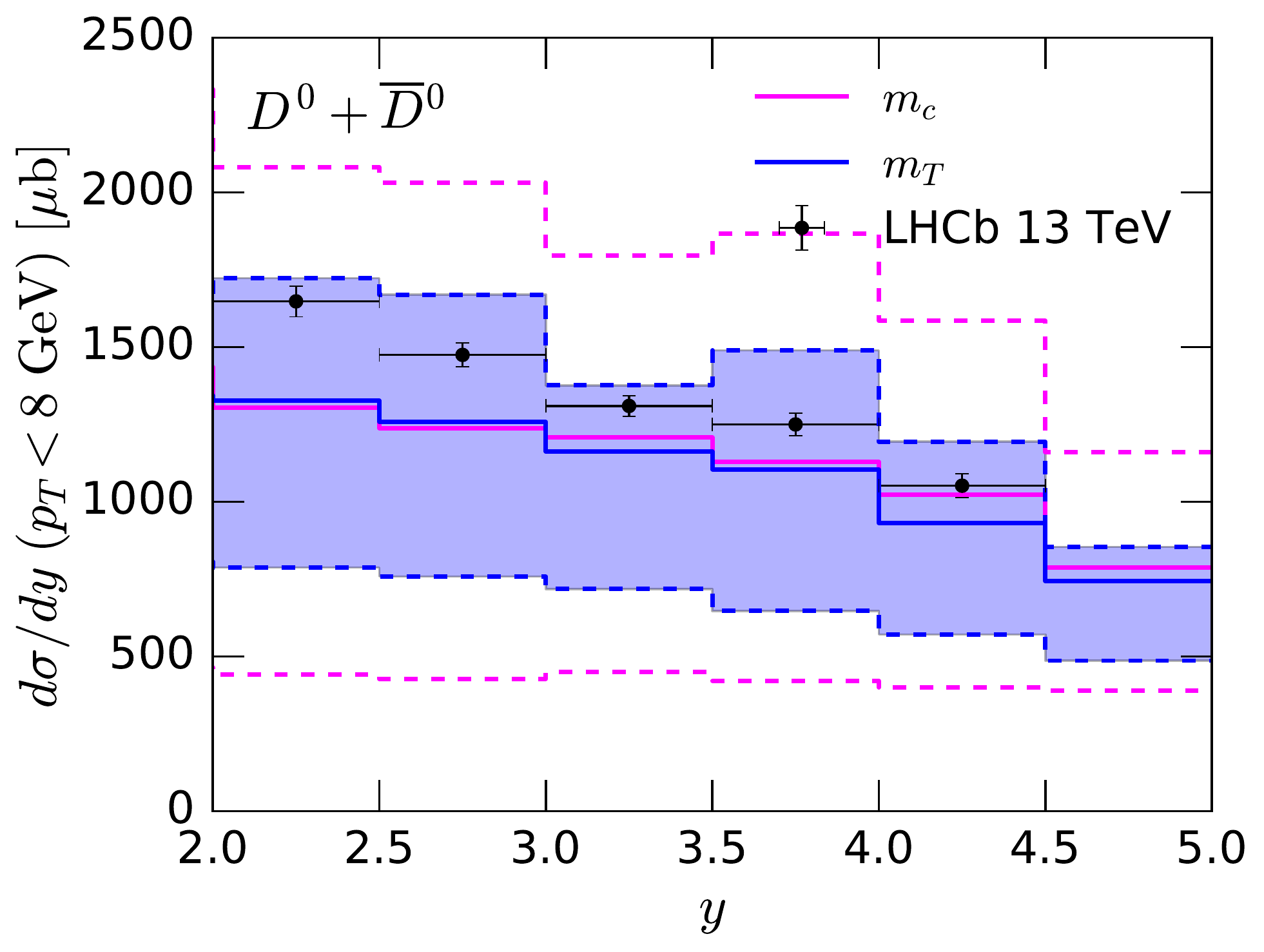}
 \includegraphics[width=4.75cm,clip]{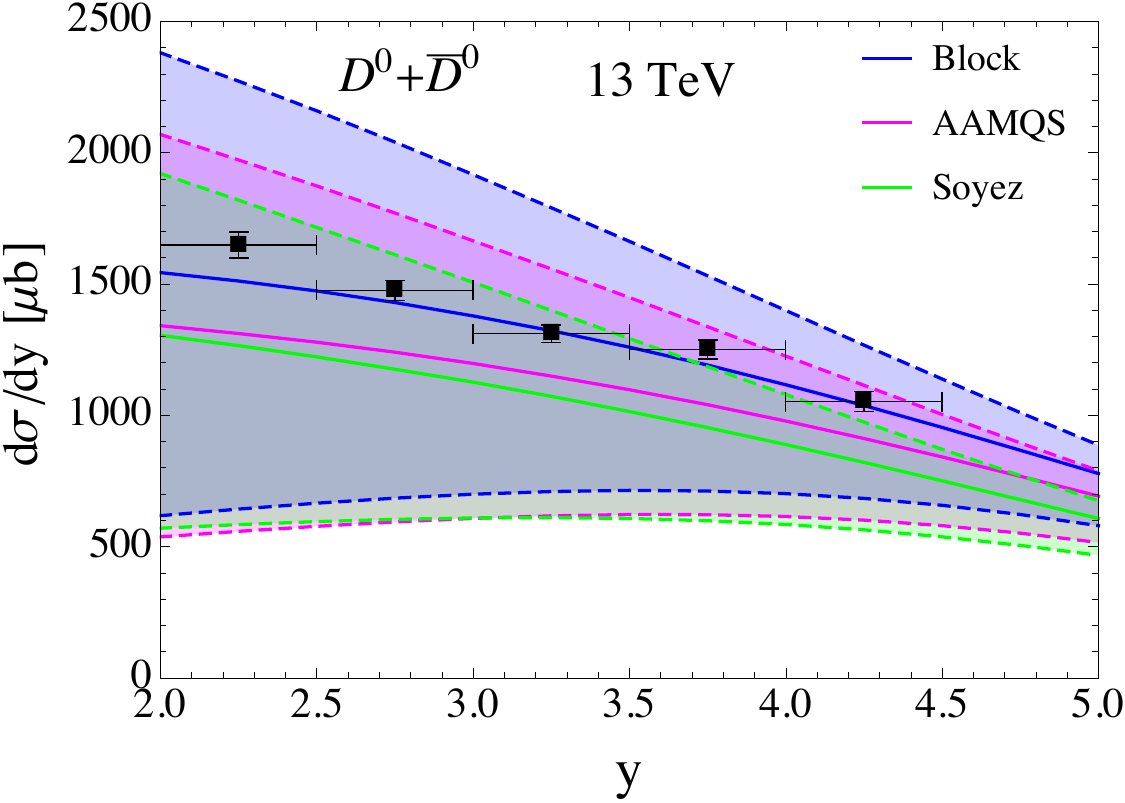}
 \includegraphics[width=4.75cm,clip]{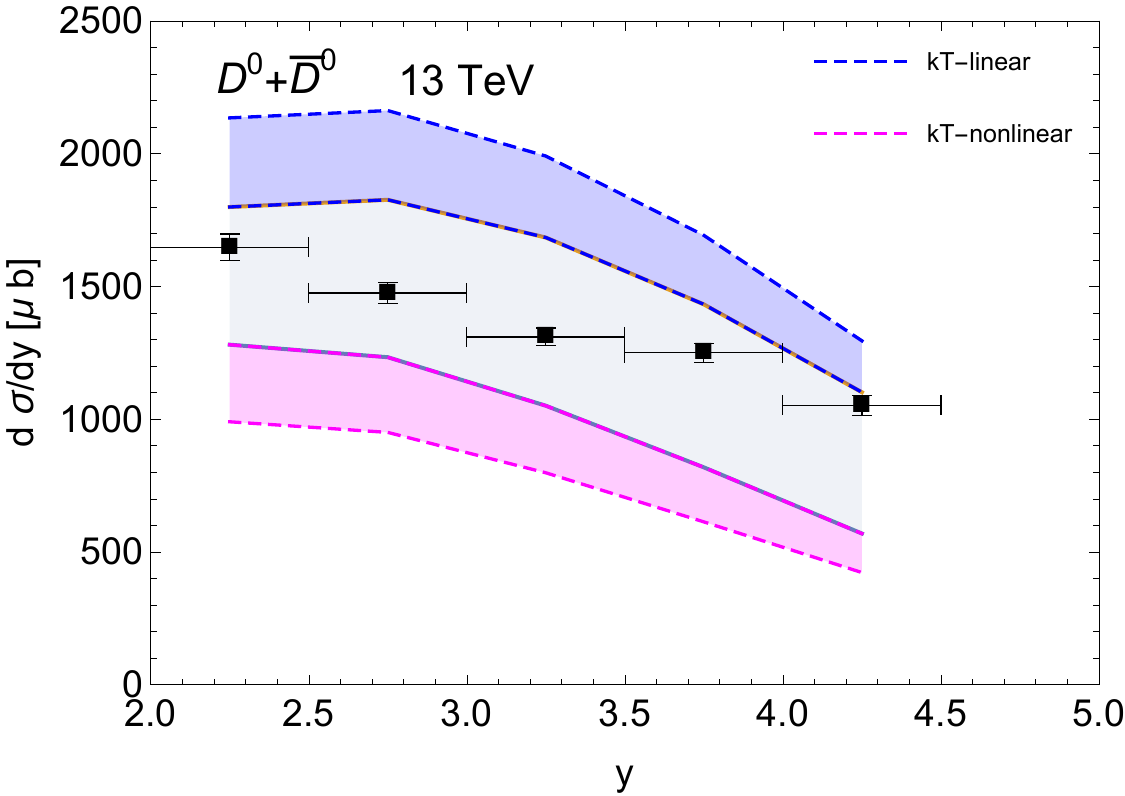}
\caption{Rapidity distribution for $pp\to D^0/\bar{D}^0\, X$
at $\sqrt{s}=7$ TeV (upper) and at $ 13$ TeV (lower) 
from NLO perturbative QCD with nCTEQ15-01 PDFs, 
three dipole models with the CT14 LO gluon PDF 
and $k_T$ factorization approach. 
Experimental data are from the LHCb measurement \cite{lhcb}. }
\label{fig-dsdy}      
\end{figure}

In this work, three different dipole cross sections are used to evaluate the prompt atmospheric neutrino flux, 
referred to as Soyez-, AAMQS- and Block-dipole, respectively. 
The Soyez dipole cross section \cite{soyez} is parameterized with the approximate form by \cite{iim}  
for the solution of the BK equation, which includes saturation \cite{bk}.
The parameters are determined by fitting to the HERA data, and this dipole result is used in the earlier evaluation in Ref. \cite{ers}. 
The AAMQS dipole presented in ref. \cite{aamqs} is improved by including the running coupling corrections to the BK equation (rcBK).  
The third dipole referred to as Block-dipole \cite{blockdm} is found approximately from the parameterization of the structure function $F_2(x,Q^2)$ \cite{block}. 

For the nuclear correction in the dipole model, we use the Glauber-Gribov method, formulated as 
\begin{equation}
\sigma_d^A(x,r)=\int d^2\vec{b}\, 2\Biggl[ 1-\exp\Biggl( -\frac{1}{2} AT_A(b)\sigma_d^p(x,r)\Biggr)\Biggr] \ , 
\end{equation}
with the impact parameter $b$. 
The nuclear profile function $T_A(b) = \int d z \rho_A (z, \vec{b})$ is normalized to unity, $\int d^2 \vec{b}\, T_A(b)= 1$.  
Here, we use a Gaussian distribution for nuclear density, $\rho_A (z, \vec{b})$.

\subsection{$k_T$ factorization}
\label{sec-kT}
Another possible approach is the so-called $k_T$ factorization scheme, which incorporates the transverse momentum of the partons.
Therefore, the cross section in this scheme is expressed in terms of the $k_T$ dependent partonic cross section and PDFs.
The cross section for this case is formulated as  
\begin{equation}
\frac{d\sigma}{d x_{F}}(s,m_Q^2) = \int \frac{dx_1}{x_1} \frac{dx_2}{x_2} \,  dz \, 
\delta(z x_1-x_F) \, x_1 g(x_1,M_F)\int \frac{dk_T^2}{k_T^2} \hat{\sigma}^{\rm off} (z,\hat{s},k_T) \, f(x_2,k_T^2) \; ,
\label{eq:ktx}
\end{equation}
with collinear approximation and the integrated distributions for incoming partons and the unintegrated PDFs, 
which incorporates the small-x resummation for partons in the target. 
This is known as the \textit {hybrid} formalism \cite{aabkl}.
In eq. (\ref{eq:ktx}), $g(x_1,M_F)$ is the integrated PDF and $f(x_2,k_T^2)$ is the unintegrated PDF. 
Parton saturation effect can be included through nonlinear evolution of the unintegrated PDF.
We investigate both cases with ($k_T$-nonlinear) or without ($k_T$-linear) saturation effects.    
Nuclear effects in $k_T$ factorization can also be included through the unintegrated parton density,
more specifically the nonlinear term in its evolution equation.
The corresponding equation is presented in ref. \cite{kutak} with detailed descriptions.

\subsection {Total and differential cross section}

Fig. \ref{fig-hqx} shows the total cross sections for the charm and bottom production from the different models 
described in the previous sections, perturbative QCD, dipole model and $k_T$ factorization.
In order to incorporate the nuclear effect, we take nitrogen for the averaged air nuclei.  
For comparison, the earlier calculation in the perturbative QCD approach in ref. \cite{berss} 
(BERSS) is also shown 
as well as the experimental data from RHIC, LHC and fixed target experiments \cite{xsection-data,lhcb}. 
At low energies, only perturbative QCD results agree well with the data. 
However, at the high energies focused in this work, all of our evaluations are consistent and have a good agreement with the data.

We also evaluate the differential cross sections in rapidity to determine the acceptable range of the scales 
comparing with the LHCb data \cite{lhcb}.
For the NLO perturbative QCD case, we take the scale variation as $(1.25, 1.48)m_{T }$ and $(4.65, 1.71)m_{T }$ 
for the lower and upper limits as guided by ref. \cite{nvf}.
In the left panels of fig. \ref{fig-dsdy}, we present the rapidity distributions for $D^0 + \bar{D}^0$ production in $pp$ collisions
for $2.0<y<4.5$ and $p_T \leq 8 \rm\ GeV$ with the corresponding data from the LHCb measurements \cite{lhcb}. 
The blue banded area presents the results evaluated with the $m_T$ dependent scales   
while the magenta dashed lines show the results with the $m_c=1.27\  \rm GeV$ dependent scales. 
The experimental data are in the narrower error bands from the scales proportional to $m_T$, 
hence we use the $m_T$ dependent scales for the uncertainties of the flux evaluations.   

Fig. \ref{fig-dsdy} also presents the predictions from the dipole models (center) and $k_T$ factorization (right). 
We have found that the differential cross sections in rapidity for $D^0 + \bar{D}^0$ production from the three different dipole models 
are consistent with experimental data at the factorization scale, $1 m_c \leq M_F \leq 4 m_c$. 
For the central values shown in the plot, we take $M_F = 2 m_c$. 
For the dipole model evaluation, since there is no explicit $p_T$ dependence
we do not apply the $p_T$ cut, which does not have the large effect in the results where the $p_T$
cut can be made. 

For the $k_T$ factorization results, the magenta and blue bands represent 
the results calculated with and without a saturation effect, respectively. 
The band width depends on the upper limit of the integration over $k_T$ in eq. (\ref{eq:ktx}), 
for which is used $2.5 m_T$ and $k_{max}$ here. 
The evaluations from the $k_T$-linear with $2.5 m_T$ and $k_T$-nonlinear with $k_{max}$
gives the gray bands in the right panels of fig. \ref{fig-dsdy}.
We use this range for the uncertainties by the scales in the prompt neutrino fluxes.

\section{Prompt atmospheric neutrino fluxes}
\subsection{Cosmic ray fluxes}
In calculating the atmospheric neutrino fluxes, the incident cosmic ray flux is one of the essential and most important factors.
In many earlier calculations, the broken power law (BPL) spectrum has been used 
with the assumption that all cosmic rays are protons.
The BPL is parameterized as 
$\phi_N(E)  = 1.7\ (E/{\rm GeV})^{-2.7}$ for $ E< 5\cdot 10^6$ GeV 
and $174 \ (E/{\rm GeV})^{-3}$ for $E>5\cdot 10^6$ GeV.
However, the recent study on the cosmic ray spectrum has provided several new parameterizations 
based on the models for the composition and sources of the cosmic rays. 
In our comprehensive work \cite{bejkrss}, we used four spectra: the BPL to compare with other predictions, 
two parameterizations from the model for three source components \cite{gaisser} 
and one spectrum for four source components (called GST) \cite{gst}.  
The two spectra in \cite{gaisser} is distinguished by the composition of cosmic rays from the extragalactic sources: 
one has a mixed composition (H3a), and the other has only protons (H3p).

In this contribution, our focus is on the effect of the different models to the prompt fluxes,
therefore it would be sufficient to compare the results with only one kind of spectrum, H3p as shown in fig. \ref{fig-fluxh3p}. 
  
\subsection{Prompt muon neutrino flux}
The atmospheric neutrino flux can be evaluated from the cascade equations which describe the propagation of high energy particles in the atmosphere.
The coupled cascade equations for protons, hadrons and neutrinos can be solved using the approximate Z-moment method.
The Z moments are the rescaled generation functions with only energy dependence,
and they are associated with the production of hadrons from the interactions and their decays to neutrinos.
The solutions of the coupled cascade equations can be written in terms of the Z moments for production and decay in the high energy and low energy limits. 
The flux of neutrinos are eventually obtained by interpolating the two approximate solutions and doing the sum over hadrons. 
More details for the Z-moment method are described in ref. \cite{lipari}.
In evaluating the muon neutrino fluxes, we take $D^0$, $D^+$, $D_s$, $\Lambda_c$ ($B^0$, $B^+$, $B_s$, $\Lambda_b$) into account for the charmed (bottom) hadrons.

\begin{figure}
\centering
\sidecaption
 \includegraphics[width=7cm,clip]{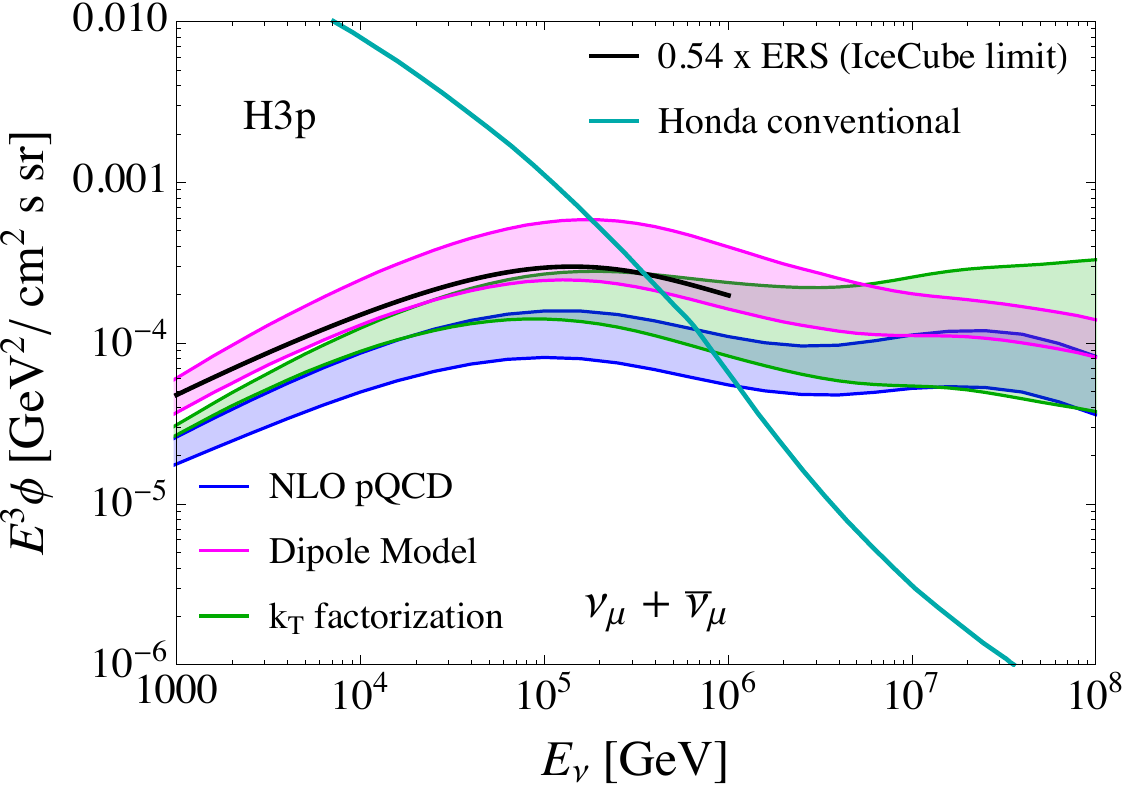}
\caption{The prompt atmospheric neutrino fluxes from the different QCD models for the H3p cosmic ray spectrum.
The upper limit from the IceCube collaboration and the conventional neutrino flux are also presented for comparison. }
\label{fig-fluxh3p}     
\end{figure}

Fig. \ref{fig-fluxh3p} shows the resulting neutrino fluxes with the H3p spectrum and all the approaches introduced in section \ref{sec-2}. 
We have found that the nuclear correction has different effect according to the models: about $20\%-30\%$ for the NLO pQCD, 
$10\%-20\%$ for the dipole models, and $30\%-50\%$ for the $k_T$ factorization at $10^5-10^8$ GeV.  
The predictions using the dipole model are larger than the fluxes from the NLO pQCD calculation all over the energy range.
For the $k_T$ factorization case, the flux with the linear evolution is the largest above $E \sim 10^7$ GeV, while that from nonlinear evolution with nuclear correction 
is close to the lower bound of the NLO pQCD evaluation. 
The upper limit from the IceCube data based on 3 year observation are also presented for comparisons \cite{icecube}.   
This IceCube limit excludes most of the dipole model results and quite constrains the upper values of the $k_T$ factorization results in this work. 
Only new predictions in the NLO pQCD and $k_T$ factorization  with nuclear corrections are placed safely below the IceCube limit.

\section {Summary}
We have investigated the prompt atmospheric neutrino flux in the different frameworks for heavy quark production, 
NLO pQCD, three kinds of the dipole models and $k_T$ factorization. 
We set the allowed QCD scales by comparisons of the total cross sections for heavy quark production 
with the experimental data from RHIC and LHC \cite{nvf}.
We checked that the differential cross sections for $D^0/\bar{D}^0$ production with these scales are consistent with the LHCb data.
The overall prompt neutrino flux is lower from NLO pQCD calculation than from the other two evaluations. 
The nuclear corrections are most significant in the $k_T$ factorization case based on the nonlinear evolution of the gluon density.
In this work, we find that the nuclear corrected NLO pQCD prediction
and the nuclear corrected $k_T$ factorization with saturation (nonlinear evolution) can safely survive from the IceCube limit based on the 3 year data.
However, a new limit was recently released based on the 6 year data by IceCube after our study \cite{icecube-six},
and we note that our prediction from all approaches are below this new limit.

\section* {Acknowledgments}
This research was supported by the US DOE, the NRF of Korea, the NSC of Poland, 
the SRC of Sweden and the FNRS of Belgium.

%
%
%

\end{document}